\documentclass[twocolumn,showpacs,preprintnumbers,amsmath,amssymb,superscriptaddress]{revtex4}
\usepackage{graphics}
\usepackage{graphicx}
\usepackage{epsfig}
\usepackage{amssymb}
\usepackage{multirow}
\usepackage{slashbox}
\begin{document}

\title{Quantum key distribution between two groups using secret sharing}

\author{S. Choi}
\affiliation{Basic Research Laboratory, Electronics and
Telecommunications Research Institute, Daejeon 305-350, Korea}
\author{J. Kim}
\affiliation{School of Electrical Engineering and Computer
Science, Seoul National University, Seoul 151-744, Korea}
\author{D. P. Chi}
\affiliation{School of Mathematical Sciences, Seoul National
University, Seoul 151-742, Korea}
\date{\today}

\begin{abstract}
In this paper, we investigate properties of some multi-particle
entangled states and, from the properties applying the secret
sharing present a new type of quantum key distribution protocols
as generalization of quantum key distribution between two persons.
In the protocols each group can retrieve the secure key string,
only if all members in each group should cooperate with one
another. We also show that the protocols are secure against an
external eavesdropper using the intercept/resend strategy.
\end{abstract}

\pacs{03.67.Dd, 03.67.Hk, 03.65.Ud}

\maketitle

\section{Introduction}
The computational power of quantum computers has threatened
classical cryptosystems. For example, public key cryptosystems,
such as Rivest-Shamir-Adleman public key cryptosystem \cite{RSA},
can be broken by quantum computers to be able to perform the fast
factorization. On the other hand, quantum mechanical phenomena
provide us a new kind of cryptosystems, called quantum key
distribution (QKD), from which we can in principle obtain
perfectly random and secure key strings.

The first quantum cryptographic protocol was presented by Bennett
and Brassard \cite{BB} and their protocol bore the acronym BB84.
In 1991, Ekert \cite{EK} proposed a QKD protocol using entangled
particles. It was modified by Bennett, Brassard, and Mermin
\cite{bbm}. Let us call the modified version the
Einstein-Podolsky-Rosen (EPR) protocol. The EPR protocol is a QKD
between two persons using an EPR pair of spin $\frac{1}{2}$
particles in the state $\frac{1}{\sqrt{2}}({\left| {00}
\right\rangle}+{\left| {11} \right\rangle})$.

Using the Greenberger-Horne-Zeilinger (GHZ) state
$\frac{1}{\sqrt{2}}({\left| {000} \right\rangle}+{\left| {111}
\right\rangle})$ \cite{GHZ} the secret sharing protocol was
presented by Hillery, Bu$\check{\rm{z}}$ek and Berthiaume
\cite{h}. In this protocol, Alice distributes the information on a
key to Bob and Charlie. And the key can be restored only when
their information are collected by them.

In this paper, applying the secret sharing protocol, we generalize
the EPR protocol on noiseless channels by the properties of
several cat states \cite{Bose} and then obtain QKD protocols
between group $A$ and group $B$. In each group the information of
a secret key is distributed to all members. After the process for
recovery of the key, the two groups get the secret key. And the
protocols require each member's approval and cooperation.
Furthermore, when some members try to affect the shared bit
adversely, if the shared key does not have the correct correlation
(or anti-correlation) then it should be revealed to others in the
test step. Any external eavesdropper should also be detected even
if several members assist the eavesdropper.

This paper is organized as follows: In Section 2, we investigate
some properties of several cat states. The QKD protocol between
two groups and its modification are presented in Section 3. We
analyze the security for the protocol in Section 4.

\section{Nonorthogonal cat states}\label{Sec:nonorthogonal}
Let us begin with reviewing cat states \cite{Bose}. The t-particle
cat state is defined as a entangled state of the type
\begin{equation}\label{eq:cat}
 \bigotimes^t_{i=1}{\left| {u_i} \right\rangle}\pm \bigotimes^t_{i=1}{\left| {u^c_i} \right\rangle}
\end{equation}
whereby $u_{i}$ stands for the binary variable in $\{0,1\}$, and
$u_{i}^{c}=1-u_{i}$. Furthermore, Equation (\ref{eq:cat}) becomes
one of the Bell states when $t=2$ and one of the GHZ states when
$t=3$.

From now on, we use the following several cat states:
\begin{eqnarray}\label{cat state1}
{\left| {\Phi_t^\pm}
\right\rangle}&=\frac{1}{\sqrt{2}}({\bigotimes^t_{i=1}\left| {0}
\right\rangle} \pm \bigotimes^t_{i=1}{\left| {1}
\right\rangle})\\\label{cat state2}
{\left| {\Lambda_t^\pm}
\right\rangle}&= \frac{1}{\sqrt{2}}(\bigotimes^t_{i=1}{\left| {0}
\right\rangle} \pm i{\bigotimes^t_{i=1}\left|{1}\right\rangle}).
\end{eqnarray} We define
${\left| {0} \right\rangle}_{x}  = {\left|{\Phi_1^+}
\right\rangle}$, ${\left| {1} \right\rangle}_{x} =
{\left|{\Phi_1^-} \right\rangle}$, ${\left|{0} \right\rangle}_{y}
= {\left| {\Lambda_1^+} \right\rangle}$, and ${\left|{1}
\right\rangle}_{y} = {\left| {\Lambda_1^-} \right\rangle}$.

For $n=k+l$, we notice the states in Equation (\ref{cat state1})
and (\ref{cat state2}) have the following relations:
\begin{eqnarray}\label{equ:decomposition}
{\left| {\Phi_n^\pm} \right\rangle}_{AB}
 &=\frac{1}{\sqrt{2}} \left({\left| {\Phi_k^+} \right\rangle}_{A}{\left|{\Phi_l^\pm}
 \right\rangle}_{B}
 + {\left| {\Phi_k^-}  \right\rangle}_{A}{\left| {\Phi_l^\mp} \right\rangle}_{B}\right)\\
 &=\frac{1}{\sqrt{2}} \left({\left| {\Lambda_k^+} \right\rangle}_{A} {\left| {\Lambda_l^\mp} \right\rangle}_{B} +
 {\left| {\Lambda_k^-} \right\rangle}_{A}{\left| {\Lambda_l^\pm} \right\rangle}_{B}\right),\\
{\left| {\Lambda_n^\pm} \right\rangle}_{AB}
  &=\frac{1}{\sqrt{2}} \left({\left| {\Phi_k^+} \right\rangle}_{A}{\left| {\Lambda_l^\pm}
 \right\rangle}_{B}
 + {\left|{\Phi_k^-}  \right\rangle}_{A}{\left| {\Lambda_l^-}\right\rangle}_{B} \right)\\
 &=\frac{1}{\sqrt{2}}\left({\left| {\Lambda_k^+} \right\rangle}_{A}{\left| {\Phi_l^\pm} \right\rangle}_{B}
 + {\left| {\Lambda_k^-} \right\rangle}_{A}{\left| {\Phi_l^\mp}
 \right\rangle}_{B}\right).
\end{eqnarray}
When $G$ is a group of t persons, assume that, for one of the
above four cat states, each person takes its one particle and
measure in the $x$- or $y$-direction. Firstly we let
$\mathcal{N}_{\it{y}}^{G}$ be the number of members modulo 4 who
measure in the $y$-direction, $\mathcal{M}_{\it{y}}^{G} =
\left\lfloor\frac{\mathcal{N}_{\it{y}}^{G}}{2}\right\rfloor$, and
$\mathcal{P}^{G}$ the sum of the measurement outcome of all
members modulo 2. Then the following results are obtained.

\begin{itemize}
\item[(a)] Suppose $\mathcal{N}_{\it{y}}^{G}$ is even. Then
$\mathcal{P}^{G}\oplus \mathcal{M}_{\it{y}}^{G}$ is $0$ for
${\left| {\Phi_{t}^{+}} \right\rangle}$, and
 it is $1$ for ${\left| {\Phi_{t}^{-}} \right\rangle}$, where
 $a \oplus b \equiv a+b$ (mod 2) for any $a,b\in\mathbb N$.

\item[(b)]Suppose $\mathcal{N}_{\it{y}}^{G}$ is odd. Then
$\mathcal{P}^{G}\oplus \mathcal{M}_{\it{y}}^{G}$ is $0$ for
${\left| {\Lambda_{t}^{+}} \right\rangle}$, and it is $1$ for
${\left|{\Lambda_{t}^{-}} \right\rangle}$.
\end{itemize}
Also, it is noticed that if the above suppositions of
$\mathcal{N}_{\it{y}}^{G}$ are not satisfied,
$\mathcal{P}^{G}\oplus \mathcal{M}_{\it{y}}^{G}$ becomes 0 or 1
with probability $\frac{1}{2}$ i.e. it has no rules.

Using an induction on $t$ the proof of such facts is given. To
begin with, for $t=1$ it is trivial. Assume that these statements
are true for $t-1$. The cat state ${\left| {\Phi_{t}^{+}}
\right\rangle}$ is considered. Let $\mathcal{N}_{\it{y}}^{G}$ be
even. Equation (\ref{equ:decomposition}) implies

\begin{equation}\label{equ:catPhi}
{\left|{\Phi_{t}^{+}}
\right\rangle}=\frac{1}{\sqrt{2}}\left({\left|{0}
\right\rangle}_{x} {\left|{\Phi_{t-1}^{+}} \right\rangle} +
{\left|{1} \right\rangle}_{x}{\left| {\Phi_{t-1}^{-}}
\right\rangle}\right).
\end{equation}

If any one member takes measurement in the $x$-direction and
obtains 0 then $\mathcal{N}_{\it{y}}^{G} =
\mathcal{N}_{\it{y}}^{G'}$ and $\mathcal{N}_{\it{y}}^{G'}$ will be
even, where G$'$ is the group of all members except that member.
From Equation (\ref{equ:catPhi}) $\mathcal{M}_{\it{y}}^{G'} \oplus
\mathcal{P}^{G'}\equiv 0$ and $\mathcal{P}^{G}=\mathcal{P}^{G'}$.
Thus $\mathcal{M}_{\it{y}}^{G} \oplus \mathcal{P}^{G}\equiv 0$.
Otherwise, $\mathcal{M}_{\it{y}}^{G'} \oplus
\mathcal{P}^{G'}\equiv 1$ by (\ref{equ:catPhi}) and
$\mathcal{P}^{G}\equiv \mathcal{P}^{G'}\oplus 1$. Thus
$\mathcal{M}_{\it{y}}^{G} \oplus \mathcal{P}^{G}\equiv 0$.

On the other hand, for the case that the member takes a
measurement in the $y$-direction, the proof is similar to the
above case. Hence, we hold that $\mathcal{M}_{\it{y}}^{G} \oplus
\mathcal{P}^{G}\equiv 0$. That is, the previous assumption holds
for $t$.
For other cat states, all of the proofs are similar.

Now, we consider two parties, $A$ and $B$, that consist of $k$
members and $l$ members, respectively. Applying the previously
described properties of the cat states, we obtain the Table
\ref{Table:Exencoding}.

\begin{table*}[htb]
\begin{center}\caption{ Relations between outcomes of $A$ and $B$.}
\begin{tabular}[t]{c|c|cc|cc}
& & \multicolumn{2}{c}{A} & \multicolumn{2}{c}{B}
\\\hline
  &  $\mathcal{N_{\it{y}}^\mathrm{A}} + \mathcal{N_{\it{y}}^\mathrm{B}}$ &
$\mathcal{N_{\it{y}}^\mathrm{A}}$ &
$\mathcal{M}_{\it{y}}^\mathrm{A}+\mathcal{P}^\mathrm{A}$ &
$\mathcal{N_{\it{y}}^\mathrm{B}}$ &
$\mathcal{M}_{\it{y}}^\mathrm{B}+\mathcal{P}^\mathrm{B}$\\
\hline
 \multirow{4}{18mm}{${\left| {\Phi_n^+} \right\rangle}$(${\left| {\Phi_n^-} \right\rangle}$)}&\multirow{4}{8mm}{even}
 &\multirow{2}{4mm}{even} & 0 & \multirow{2}{4mm}{even}  & 0(1) \\
 &  &  & 1 &   & 1(0)\\
 \cline{3-6}
 & &\multirow{2}{4mm}{odd} & 0 & \multirow{2}{4mm}{odd}  & 1(0)\\
 &     &      & 1   &                & 0(1)\\\hline
 \multirow{4}{18mm}{${\left| {\Lambda_n^+}
\right\rangle}$(${\left| {\Lambda_n^-}
\right\rangle}$)}&\multirow{4}{4mm}{odd}
 &\multirow{2}{4mm}{even} & 0 & \multirow{2}{4mm}{odd}  & 0(1) \\
 &  &  & 1 &   & 1(0)\\
 \cline{3-6}
 & &\multirow{2}{4mm}{odd} & 0 & \multirow{2}{4mm}{even}  & 0(1) \\
 &     &      & 1   &                & 1(0)\\\hline
 \end{tabular}\label{Table:Exencoding}
\end{center}
\end{table*}

\section{Protocols}\label{Sec:QKD1}
By means of the properties of the cat states, we describe the QKD
protocols between two groups. We first discuss how two groups
proceed to share the secret key string. Next, by modifying several
steps we show to be able to use the cat states efficiently.


\subsection{Protocol}\label{section1}
In this section, we present a QKD protocol between two groups, $A$
and $B$, that consist $k$ ($k>1$) members and $l$ ($l> 2$) members
respectively. From here, with $n=k+l$ we use the $n$-particle cat
states and suppose that all members are arbitrarily ordered . For
each shared bit, each group requires a member who collects the
information that has been distributed to all members. We call such
members the `collectors'. We present one of the methods to collect
the information after description of the protocol. We presume that
a collector chooses the used cat state and its information is
possessed by only the collector. However, if secure classical
channels among members in $A$ exists, all members in $A$ may share
the information on demand and then may choose the cat state
together under their agreement.

\begin{enumerate}
\item[1.] \label{pro1:select}A collector in $A$ randomly chooses
an $n$-particle cat state out of ${\left|{\Phi_{n}^{\pm}}
\right\rangle}$ and ${\left|{\Lambda_{n}^{\pm}}  \right\rangle}$
which is denoted by ${\left| {S} \right\rangle}$. Each particle of
${\left|{S}  \right\rangle}$ is transmitted to each member of the
two groups.

\item[2.] \label{pro1:meas}Each member of the two groups randomly
performs a measurement on his own particle either in the $x$- or
$y$-direction, respectively.

\item[3.] \label{pro1:annbasis} Each member in the two groups
announces the  basis he used through the public channel, but not
the result he obtained, . The two groups, $A$ and $B$, obtain
$\mathcal{N}_{\it{y}}^\mathrm{A}$ and
$\mathcal{N}_{\it{y}}^\mathrm{B}$, respectively. We call the
member who finally announces the basis in each group the `last
member' Here, the announcement of the last member in $A$ should be
followed by $B$'s.

\item[4.] \label{pro1:parity}Two groups, $A$ and $B$, collect the
outcomes to obtain $\mathcal{P}^\mathrm{A}$ and
$\mathcal{P}^\mathrm{B}$, respectively, and then obtain the shared
bit $\mathcal{M_{\it{y}}^\mathrm{A}} \oplus\mathcal{P}^\mathrm{A}$
and $\mathcal{M_{\it{y}}^\mathrm{B}}\oplus
\mathcal{P}^\mathrm{B}$, respectively. The last member is never
the collector and it will be discussed in Section
\ref{sec:conspirator}.

In order to obtain the key bit strings, the two groups should
repeat the above steps a sufficient number of times.

\item[5.] \label{pro1:test}The two groups have a public discussion
on a set of bits used to detect an eavesdropper's presence. For
the test bits, $A$ reveals $\mathcal{P}^\mathrm{A}$ and is
followed by B. The reason will be treated in Section
\ref{sec:conspirator}.

\item[6.] \label{pro1:annSource}$A$ announces the cat states
${\left| {S} \right\rangle}$ that were chosen at first. For
${\left|{S} \right\rangle}= {\left|{\Phi_{n}^{\pm}}
\right\rangle}$, if $\mathcal{N}^{\mathrm{A}}_{\it{y}}+
      \mathcal{N}_{\it{y}}^{\mathrm{B}}$ is
even, then the shared bit will be kept, and otherwise, it will be
discarded. In case ${\left|{S} \right\rangle}
={\left|{\Lambda_{n}^{\pm}} \right\rangle}$, if
$\mathcal{N}^{\mathrm{A}}_{\it{y}}+
\mathcal{N}_{\it{y}}^{\mathrm{B}}$ is odd, it will be kept, and
otherwise, it will be discarded. So the two groups keep it with
probability $\frac{1}{2}$.

With a set of test bits, the two groups make independently a test
to detect the presence of eavesdroppers or the faulty bit string
made by some members who behave wrong.

If an error exists, all shared keys should be discarded, and the
two groups should go back to Step 1. Otherwise, they go on the
next step.
\end{enumerate}

\begin{figure}[t]\caption{Obtaining
$\mathcal{P}=\mathcal{P}^{\mathrm{A}}$ or
$\mathcal{P}^{\mathrm{B}}$: $E_{i}$ is the outcome of the $i$-th
member, $R$ is the random bit chosen by the collector, and $m$ is
the number of all members.}\label{particle}
\begin{center}
\mbox{\epsfig{figure=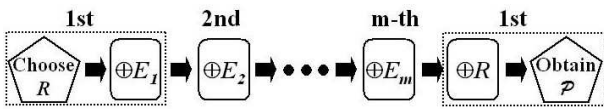, height=1.5cm}}
\end{center}
\end{figure}

We suggest a method of obtaining $\mathcal{P}^{\mathrm{A}}$ (or
$\mathcal{P}^{\mathrm{B}}$). Here, we consider the first member as
a collector and all operations are module 2. The collector chooses
a random bit `$R$', adds it to his outcome, and sends the result
to the second member. The second member adds his own outcome to
the received one, and then gives it to the next member. This
procedure is continued until the collector receives
$\mathcal{P}^{\mathrm{A}} ($or $ \mathcal{P}^{\mathrm{B}}) \oplus
R$. After that, the collector finally takes
$\mathcal{P}^{\mathrm{A}} ($or $ \mathcal{P}^{\mathrm{B}})$, which
is $\mathcal{P}^{\mathrm{A}} ($or $\mathcal{P}^{\mathrm{B}})
\oplus R\oplus R$ (see Figure \ref{particle}).

If each member plays a role of the collector in rotation, all
secret key string should be divided among all members with the
same portion. If without rotation just one member always plays the
collector, the protocol may be similar to the EPR protocol.
However, even in such a case it is not the same as the EPR
protocol in the aspect of requiring all members' approval. For
instance, a message from another group is never decrypted without
all members' agreement.

\subsection{Modified protocol with a chairperson}\label{section2}
In the previous protocol, each shared bit is discarded with
probability $\frac{1}{2}$ in Step 6. By modifying a method to
perform the measurement in the previous protocol, the cat states
can efficiently be used, i.e., the number of shared bits which are
discarded can be decreased. In here, we assume that $k>1$ and
$l>1$. In the modified protocol, a specific member, called the
`chairperson', keeps his own qubit intact until the other members
announce their information on the bases, and then  takes his own
basis dependent on ${\left| {S} \right\rangle}$ to prohibit the
bit from being discarded. Because the chairperson needs
information on ${\left|{S} \right\rangle}$, a collector in $A$
should play the chairperson. We clearly remark that any member can
play the chairperson if all members in $A$ have the information on
${\left|{S} \right\rangle}$.

To obtain a more efficient protocol, Step 3 and 4 in the previous
protocol are modified as the followings.

\begin{enumerate}
\item[3$'$.] \label{pro2:measurement}Except the chairperson each
member in the two groups randomly performs a measurement on his
own particle either in $x$- or $y$-direction.

\item[4$'$.] \label{pro2:annbasis}
\begin{enumerate}
\item Let $A^{\prime}$ be the group consisting of all members in
$A$ except the chairperson. All members in $A^{\prime}$ and $B$
announce the measurement bases. Then the two groups get
$\mathcal{N}_{\it{y}}^\mathrm{A^{\prime}}$ and
$\mathcal{N}_{\it{y}}^\mathrm{B}$, respectively. Now, the
collector in $B$ never plays the last member.

\item \label{Pro2:annmeas} Using the properties of cat states, the
chairperson performs the measurement on his particle depending on
$\mathcal{N}_{\it{y}}^\mathrm{A^{\prime}} +
\mathcal{N}_{\it{y}}^\mathrm{B}$ and ${\left| {S} \right\rangle}$
in order to prevent the shared bit from being discarded.
\end{enumerate}
\end{enumerate}

We remark that the order of the basis announcements of two groups
is not important in this protocol because
$\mathcal{N}_{\it{y}}^\mathrm{A}$ is determined by
$\mathcal{N}_{\it{y}}^\mathrm{B}$ and ${\left| {S}\right\rangle}$.

\section{Analysis of Security}\label{QKD1:security} In this section, we
analyze security of our protocols. Firstly, we discuss the case
that several members have some wrong behavior. The second case
treat an eavesdropper who uses the intercept/resend strategy
\cite{bbbss}. We again divide the second case into two cases
according to the existence of members who give an eavesdropper
some helps. Since the protocols should be secure even if all
members in $A$ share the information on ${\left|{S}
\right\rangle}$ before the transmission of the particles in the
first step, we assume that all members in $A$ know the
information.

\subsection{Members with wrong behaviors}\label{dishonest}
In this part, we discuss that when there are some members who
behave wrong if two groups, particularly the collectors, have the
faulty key strings then they can notice it from the test.

To begin with, we treat a chairperson in the modified protocol.
The measurement basis of the chairperson is exactly determined by
the other members' ones and the state ${\left|{S}\right\rangle}$.
Thus, he cannot change his basis arbitrarily, and can affect only
his measurement result. From the above fact, we clearly obtain
that he can have no more influence on the key information than the
other members' one. Therefore, it suffices to consider the
investigation of the other members' behavior.

We think over all members' behavior except the collector's one.
Because $\mathcal{P}^\mathrm{A}$ (or $\mathcal{P}^\mathrm{B}$) is
possessed by just a collector, any member except the collector
cannot know it and hence cannot notice the shared bit. While some
members are having behavior wrong, they cannot perceive what is
the key bit made by their actions. Moreover, before the test step
they cannot perceive if errors will be detected in the test step
and what are the bit stings used to test. Hence, the nonexistence
of errors in sufficiently many test bits implies that almost all
the key bits have correct correlations (or anti-correlations).
Since the collectors only possess the shared bits, if the bits
have the correct correlations (or anti-correlations) the two
groups can share correct keys although there exist some members to
behave wrong. Therefore, if some members have a wrong effect on
some of bit strings and if the two groups share the faulty bit
strings, then they can find errors from sufficiently many test
bits.

\subsection{Eavesdropper and conspirators in two
groups}\label{sec:conspirator} Suppose that there are an
eavesdropper and some members who assist her. Here, the
eavesdropper and the members are called `Eve' and `conspirators',
respectively. We first discuss that without assistance of
conspirators Eve uses the intercept/resend strategy \cite{bbbss},
and then discuss that with some helps of conspirators Eve uses
such strategy and the $l$-particle entangled state to resend to
$B$. Finally it is treated that under the same strategy Eve uses
the $n'$-particle entangled state to resend to $B$ ($n'> l$).

\subsubsection{No conspirator in intercept/resend strategy}
We consider that Eve uses the intercept/resend strategy, i.e., Eve
intercepts $l$ particles travelling from $A$ to $B$, performs a
measurement on that particles, and resends  an $l$-particle fake
state instead. Even if Eve chooses a fake state according to the
measurement result and resends it, the two groups will detect an
error in the shared bit with probability $\frac{1}{8}$ for the
original protocol, and with probability $\frac{1}{4}$ for the
modified protocol, respectively. The difference of probability
comes from the nonexistence of the discarded bits in the modified
protocol. Hence, the two groups can find errors for sufficiently
many test bits.

\subsubsection{Using intercept/resend strategy: Eve and
conspirators}

We consider that Eve adopts the intercept/resend strategy and has
some conspirators in two groups. First conspirators should try to
change $\mathcal{N_{\it{y}}^\mathrm{A}}$,
$\mathcal{N_{\it{y}}^\mathrm{B}}$, $\mathcal{P}^\mathrm{A}$ or
$\mathcal{P}^\mathrm{B}$, to make no errors which are caused by
Eve's eavesdropping. However, as stated in Section
\ref{dishonest}, it is impossible for any member except the
collector to change $\mathcal{P}^\mathrm{A}$ and
$\mathcal{P}^\mathrm{B}$ into what they want. In addition, any
member except the last members can never change
$\mathcal{N_{\it{y}}^\mathrm{A}}$ and
$\mathcal{N_{\it{y}}^\mathrm{B}}$ into what they want. Thus, it
suffices to treat the case that one more conspirator plays
collectors (or last members) under assumption that the collector
(or the last members) is played in rotation by each member.

We assume that Eve eavesdrops with the probability $\lambda$,
$0\leq \lambda \leq 1$ using the intercept/resend strategy;
$\lambda =0$ means that Eve is not eavesdropping at all. Let
$r_{a}$ be the number of conspirators in $A$ and $r_{b}$ the
number of conspirators in $B$. The two groups randomly select $t$
shard bits in order to estimate the error rate. In the first
protocol, Eve's eavesdropping then causes at least the following
error rates according to $r_{a}$ and $r_{b}$.

In the case that $r_{a}=0$ and $r_{b}\geq 1$, we have
\begin{equation}\label{equ:groupApro1}
 1-\left(\frac{7}{8}\right)^{\lambda t\left(1-\frac{r_{b}-1}{l}\right)}.
\end{equation}
Applying Eve's the measurement result on the intercepted
particles, they can notice the values of
$\mathcal{N_{\it{y}}^\mathrm{B}}$ and $\mathcal{P}^\mathrm{B}$ to
make no errors. For example, in the case that the measurement
result is ${\left|{\Phi_{l}^{+ }} \right\rangle}$, if
$\mathcal{N_{\it{y}}^\mathrm{B}}$ is even and
$\mathcal{M_{\it{y}}^\mathrm{B}}+\mathcal{P}^\mathrm{B}$ is 0 then
there is no error. So, only having assistance of a collector and a
last member in $B$ at once, they can forbid errors to be caused.
They have the chance with ratio $\frac{r_{b}-1}{l}$ for one shared
bit, i.e., the conspirators in $B$ can play a collector and a last
member simultaneously with the ratio. Hence Equation
(\ref{equ:groupApro1}) is obtained.

We note that if a last member is a collector in $B$ then the ratio
becomes greater than $\frac{r_{b}-1}{l}$. Thus, in the first
protocol, a last member can never be the identical person with a
collector during one key agreement.

In the case $r_{a}\geq 1$ and $r_{b} =0$, we have
\begin{equation}\label{equ:groupBpro1}
 1-\left(\frac{7}{8}\right)^{\lambda t\left(1-\frac{r_{a}-1}{k}\right)}.
\end{equation}
From the information on ${\left|{S} \right\rangle}$ and Eve's
measurement result they can find the suitable values for
$\mathcal{N_{\it{y}}^\mathrm{A}}$ and $\mathcal{P}^\mathrm{A}$
which induce no error. Thus Eve and conspirators in $A$ can change
these values into the found suitable ones, only if the
conspirators play the collector and the last member in $A$
simultaneously. Hence Equation (\ref{equ:groupBpro1}) is found. It
also becomes the reason that a last member in $A$ never plays a
collector.

In the case $r_{a}\geq 1$ and $r_{b}\geq 1$, we have
\begin{equation}\label{equ:two groupspro1}
1-\left(\frac{7}{8}\right)^{\lambda
t\left[\left(1-\frac{r_{a}-1}{k}\right)\left(1-\frac{r_{b}+1}{l}\right)\right]}.
\end{equation}

In this case, it is clear that they are able to use two methods
discussed in the above paragraphs. In addition, they are able to
change $\mathcal{N_{\it{y}}^\mathrm{B}}$ to make the shared bit be
discarded, or change $\mathcal{P}^\mathrm{B}$ to make no error by
means of information on $\mathcal{N_{\it{y}}^\mathrm{A}}$,
$\mathcal{N_{\it{y}}^\mathrm{B}}$, $\mathcal{P}^\mathrm{A}$ and
${\left| {S} \right\rangle}$. To do so, they need assistance of
any conspirator in $A$ for ${\left| {S} \right\rangle}$ and either
the collector or the last member in $B$. The rate that such cases
occur in $B$ is $\frac{r_{b}+1}{l}$. Therefore, we obtain the
Equation (\ref{equ:two groupspro1}).

Furthermore, we perceive that $\mathcal{N_{\it{y}}^\mathrm{A}}$
have to be announced before $\mathcal{N_{\it{y}}^\mathrm{B}}$.
This is because if not, Eve is able to make no errors even with
assistance of either the collector or the last member in $A$
without any conspirator in $B$.

Now, we can notice that the probability in Equations
(\ref{equ:groupApro1}) and (\ref{equ:groupBpro1}) are not less
than in Equation (\ref{equ:two groupspro1}). So it is sufficient
to treat only Equation (\ref{equ:two groupspro1}).

For case of $r_{b}=l-1$, the probability in Equation (\ref{equ:two
groupspro1}) is 0 and then this protocol is not secure, but it is
not so in the modified protocol which will be treated later. Next,
we consider the case, $r_{a}=k-1$ and $r_{b}=l-2$.

\begin{equation}
1-\left(\frac{7}{8}\right)^{\lambda
t\left[\left(1-\frac{r_{a}-1}{k}\right)\left(1-\frac{r_{b}+1}{l}\right)\right]}\geq
0.95.
\end{equation}
if and only if
\begin{equation}
\left(1-\frac{r_{a}-1}{k}\right)\left(1-\frac{r_{b}+1}{l}\right) =
\frac{2}{k\cdot l}\lambda t \geq 270.
\end{equation}

Though $r_{a}=k-1$ and $r_{a}=l-2$, their existence can be
detected with probability 0.95 by choosing sufficiently many test
bits which satisfy $\lambda t \geq 135 k l$ . From the equations
we can know that by making test bits be increased, even for
extreme cases, eavesdropping can also be detected with as high
probability as the two groups need. However, the more many test
bits are required to detect Eve's eavesdropping as the rate of
existence of the conspirators increases.

We remark that quite many test bits should be chosen in the case
that $r_{a}=k-1$ and $r_{b}=l-2$. Hence, upon all members'
deliberation for presumption of the number of members that can
behave wrong, the number of test bits can effectively be
modulated.

\begin{figure}[h]\caption{The probability in Equation (\ref{equ:two groupspro1})
when $k=6,l=4$ and $r_{a}=3$.}\label{figure:Pro1,6,4}
\begin{center}
\mbox{\epsfig{figure=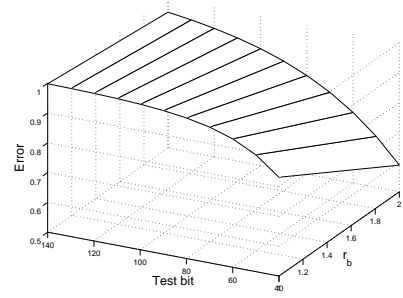, height=4cm}}
\end{center}
\end{figure}

\begin{figure}[h]\caption{The probability in Equation (\ref{equ:two groupspro1})
when $k=4,l=6$ and $r_{a}=2$.}\label{figure:Pro1,4,6}
\begin{center}
\mbox{\epsfig{figure=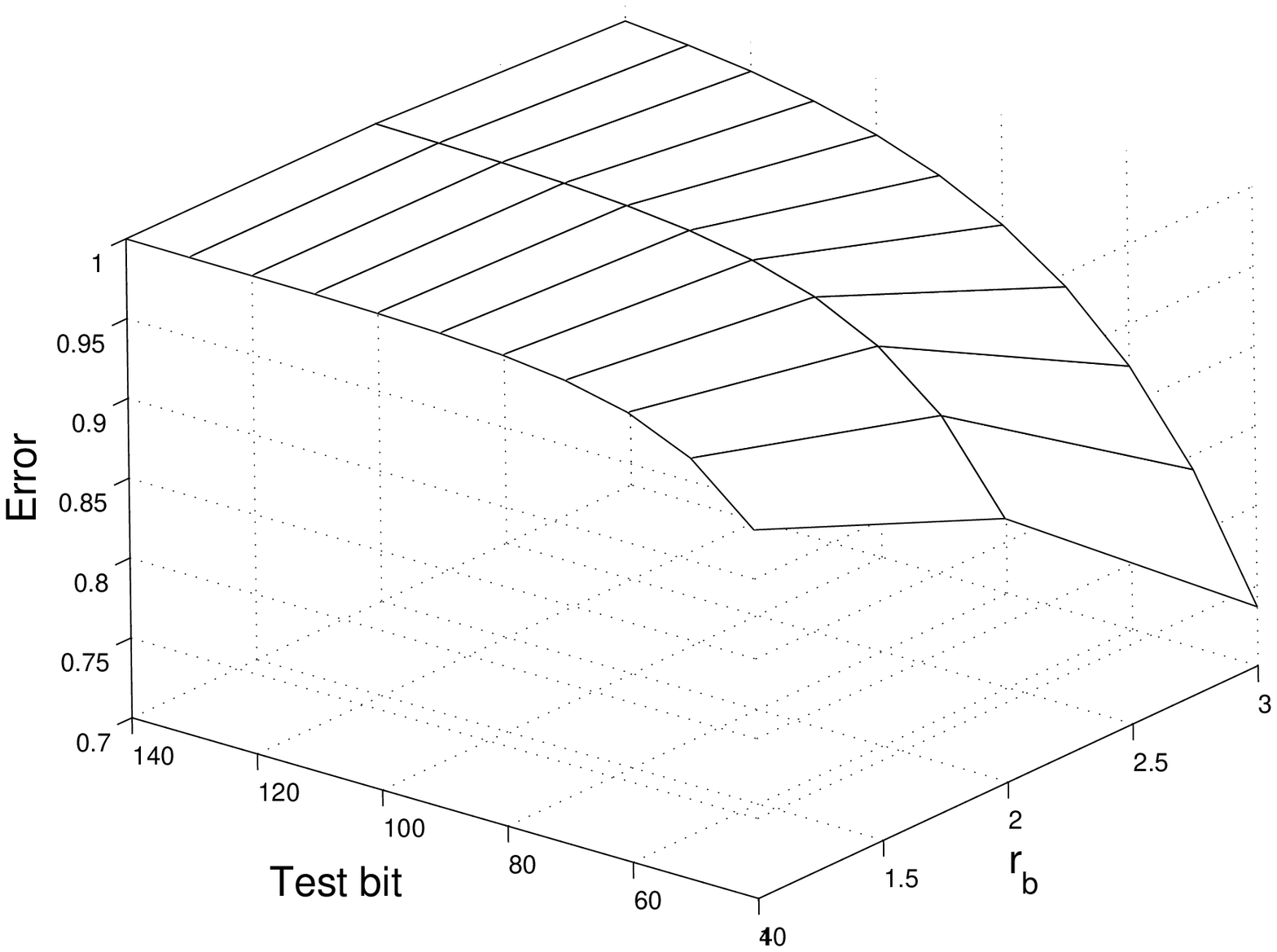, height=4cm}}
\end{center}
\end{figure}

\begin{figure}[h]\caption{The probability in Equation (\ref{equ:two groupspro1})
when $k=6, l=6$ and $r_{a}=3$.}\label{figure:Pro1,6,6}
\begin{center}
\mbox{\epsfig{figure=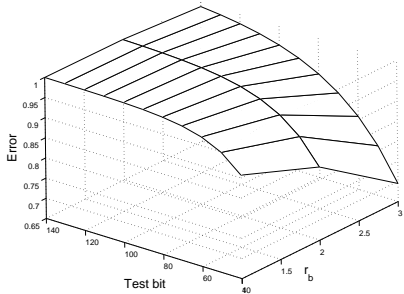, height=4cm}}
\end{center}
\end{figure}

We now consider the case that $r_{a}$ and $r_{b}$ are not more
than a half of the number of all members in $A$ and $B$,
respectively. The probability in Equation (\ref{equ:two
groupspro1}) is larger than 0.95 if and only if
\begin{equation}
\lambda
t\left(1-\frac{r_{a}-1}{k}\right)\left(1-\frac{r_{b}+1}{l}\right)
= \frac{2}{k\cdot l}\geq 22.44
\end{equation}
Then if $\frac{r_a}{k}$ and $\frac{r_b}{l}$ are fixed, the
probability increases as $k$'s value increases or $l$'s one
decreases. From comparisons between Figures \ref{figure:Pro1,6,4}
and \ref{figure:Pro1,6,6}, and between Figures
\ref{figure:Pro1,4,6} and \ref{figure:Pro1,6,6}, we can certainly
perceive the above facts.

\begin{table}[h]\caption{The probability in Equation (\ref{equ:two groupspro1})
when $k=6$, $l=6$ and $r_{a}=3$}\label{Table:pro1,6,6}
\begin{ruledtabular}
\begin{tabular}[t]{c|c|c|c|c|c|c|c}
\backslashbox{$r_b$}{$t$}
 & 20 & 40 & 60 & 80 & 100 & 120 & 140\\\hline
1  & 0.6948 & 0.9069& 0.9716 & 0.9913 & 0.9974 &  0.9992 & 0.9998\\
2  & 0.5894  & 0.8314 & 0.9308 &  0.9716  &  0.9883  &  0.9952& 0.998\\
3  & 0.4476 & 0.6948  & 0.8314  &0.9069 & 0.9486 & 0.9716 & 0.9843\\
\end{tabular}
\end{ruledtabular}
\end{table}

If $\frac{r_a}{k}\leq \frac{1}{2}, \frac{r_b}{l}\leq\frac{1}{2}$
and $\lambda =1$, it follows from the condition $l >2$ that the
number of required test bits is not less than 270. However, for
fixed $k$ and $l$ the fewer number of  test bits are required. The
change of the error rate according to the number of test bits is
exemplified in Tables \ref{Table:pro1,6,6} for the case that
$\frac{r_a}{k}\leq \frac{1}{2}$ and $\frac{r_b}{l}\leq
\frac{1}{2}$

The probability in the modified protocol has a little difference
from the primary protocol because there are no discarded bits and
$\mathcal{N_{\it{y}}^\mathrm{A}}$ is determined by
$\mathcal{N_{\it{y}}^\mathrm{B}}$ and ${\left|{S} \right\rangle}$.
By removing the strategy to use $\mathcal{N_{\it{y}}^\mathrm{A}}$
we can easily get the following error rates of the modified
protocol.

In the case $r_{a}=0$,
\begin{equation}
1 -
\left(\frac{3}{4}\right)^{\left(1-\frac{r_{b}-1}{l}\right)\lambda
t}.
\end{equation}

In the case $r_{a}\geq 1$,
\begin{equation}\label{Eq:pro2AB}
1 -
\left(\frac{3}{4}\right)^{\left(1-\frac{r_{b}}{l}\right)\lambda
t}.
\end{equation}
As in the case of the first protocol, we analyze just the case
$r_{a}\geq 1$. The error rate in Equation (\ref{Eq:pro2AB}) has no
connections with the values of $r_a$ and $k$, and depends just on
$r_b$ and $l$. For $r_b =l-1$, we require only $t$ that satisfies
$\lambda t \geq 10.4l$. For $\frac{r_b}{l}=\frac{1}{2}$, it
becomes $\lambda t \geq 21.4$. In the modified protocol, we can
notice that two groups require remarkably smaller test bits than
the first protocol, and furthermore errors can be detected from
the test step even in the cases $l=2$ or, $r_{a}=k-1$ and
$r_{b}=l-1$, while errors cannot be detected for the case in the
first protocol.

We remark that if just a collector can take the information on
${\left|{S}\right\rangle}$ or only one member plays the collector
for all shared bits then a fewer test bits would be required.

\subsubsection{Intercept/resend strategy using entangled states}

We assume that Eve intercepts $l$ particles travelling from $A$ to
$B$, and call this state `the intercepted state'. She chooses an
$n'$-particle cat state and resends $l$ particles of this cat
state to $B$ ($n^{\prime}>l$). We refer to the remainder
($n^{\prime}-l$)-particle state as `the remainder state'.

Before announcement of $\mathcal{N}_{\it{y}}^\mathrm{A}$ (or
$\mathcal{N}_{\it{y}}^\mathrm{B}$) the measurement of the
intercepted state (or the remainder state) cannot give her the
information on $\mathcal{P}^\mathrm{A}$ (or
$\mathcal{P}^\mathrm{B}$). So she should measure on the
intercepted state and the remainder state, after
$\mathcal{N}_{\it{y}}^\mathrm{A}$ and
$\mathcal{N}_{\it{y}}^\mathrm{B}$ are announced. Even though she
measures in the way, she should have the information on
${\left|{S}\right\rangle}$ to obtain $\mathcal{P}^\mathrm{A}$,
since $\mathcal{P}^\mathrm{A}$ is completely determined by
${\left|{S}\right\rangle}$ and $\mathcal{N}_{\it{y}}^\mathrm{A}$.
In order to take information on ${\left| {S} \right\rangle}$, she
needs any conspirator in $A$.

On the other hand, she wants to change $\mathcal{P}^\mathrm{A}$ or
$\mathcal{P}^\mathrm{B}$ into the values she desires to prevent
errors from occurring. Thus she needs collectors' assistance in
$A$ or $B$. Without any conspirators the test induces errors with
probability $\frac{1}{4}$ in the first protocol and probability
$\frac{1}{2}$ in the modified protocol, respectively. For these
facts Eve's strategy makes at least the following error rate in
the first protocol.

In the case $r_{a}=0$,
\begin{equation}\label{equ:groupB-2}
1-\left(\frac{3}{4}\right)^{\lambda t}.
\end{equation}
In the case $r_{a} \geq 1$,
\begin{equation}\label{equ:groupAB-2}
1-\left(\frac{3}{4}\right)^{\lambda
t\left(1-\frac{r_{a}}{k}\right)\left(1-\frac{r_{b}}{l}\right)} .
\end{equation}
In the modified protocol the error rates are similar to the first
protocol, because Eve cannot have a different strategy. From these
equations we can know that this strategy is not optimal to Eve.

\section{Summary}
Applying the properties of cat states and the secret sharing
\cite{h}, we proposed two generalized QKD protocols between two
groups and showed that the protocols are secure against an
external eavesdropper using the intercept/resend strategy. The
importance of these protocols is that any member in the two groups
cannot obtain the secret key strings without cooperation, that is,
the secret key strings can be obtained only under all member's
approval.

\textbf{Acknowledgments} S.C. acknowledges the support from
Ministry of Planning and Budget and thanks S.Lee for discussions.
D.P.C. acknowledges the support from Korea Research Foundation
(KRF-2000-0150DP0031).


\end{document}